\documentstyle{mn}

\def\be{\begin{eqnarray}}
\def\ee{\end{eqnarray}}
\title[Type I Superconductivity in Neutron Stars]
{Type I Superconductivity of Protons in Neutron Stars}
\author[D. M. ~Sedrakian, A. D. Sedrakian, G. F. Zharkov]
       {D. M. ~Sedrakian$^{1,2}$, 
        A. D. ~Sedrakian$^3$\thanks{Max Kade Foundation Research Fellow.}
        and G. F. ~Zharkov$^4$\\
$^1$Departement d'Astrophysique Relativiste et de Cosmologie,
CNRS,  Observatoire de Paris, 92195 Meudon, France \\
$^2$Department of Physics,  Yerevan State University, 375049 Yerevan, 
Armenia\\
$^3$Center for Radiophysics and Space Research, Cornell University, 14853 Ithaca, NY \\
$^4$P. N. Lebedev Physical Institute, Academy of Sciences, 117924 Moscow, Russia}

\date{Accepted 1997 May 19.
      Received 1997 May 12;
      in original form 1997 February 28}

\pagerange{203--207}
\volume{290}
\pubyear{1997}

\begin{document}

\maketitle

\begin{abstract}
\noindent
The magnetic structure of neutron vortices in the superfluid cores of 
neutron stars is determined assuming that the proton condensate 
forms a type I superconductor. It is shown that
the entrainment currents induced by the neutron vortex circulation
cause the proton superconductor to break into successive domains of 
normal and superconducting regions. The Gibbs free-energy is found
in the case in which the  normal domains form cylindrical 
tubes coaxial with the 
neutron vortex. The minimum of the energy functional corresponds to 
a tube radius $a\sim 0.1-0.5 ~b$, where $b$ is the outer radius of 
the neutron vortex. The magnetic field within the tube is of the 
order of $ 5 \times 10^{14}$ G. 
\end{abstract}

\begin{keywords}
MHD -- stars: neutron -- pulsars: general -- stars: rotation.
\end{keywords}

\section{Introduction}

The knowledge of the properties of neutron vortex lattice in the  
quantum liquid cores of a pulsar
(in the simplest case a mixture of neutrons, protons, and electrons,  
called further n-p-e phase) is important for the understanding of the 
coupling mechanisms between the superfluid and normal component of the star.
The problem differs from the case of vortices in a rotating 
uncharged superfluid mainly due to the fact that above the nuclear 
saturation density, $\rho_{\rm sat}\simeq 2.6\times 10^{14}$  g cm$^{-3}$, the $^3P_2$ superfluid neutron vortex lattice is embedded in a $^1S_0$ proton superconductor. 
Under these conditions, the relation between the superfluid mass current, 
$\bmath p$,
and superfluid velocities, $\bmath v$, in the hydrodynamic equations assumes a tensor 
character with respect to the isospin indecies ($1,\, 2$):
\be \label{0}
\left(\begin{array}{c} {\bmath p}_1 \\
{\bmath p}_2 \end{array}\right)  
=\left(\begin{array}{cc} \rho_{11}& \rho_{12}\\ 
\rho_{21}& \rho_{22}\end{array}\right) 
~
\left(\begin{array}{c} {\bmath v}_ 1 \\ {\bmath v}_ 2 \end{array}\right), 
\ee 
where $\rho_{ik}$ is the superfluid density tensor, which replaces the 
scalar superfluid density in the Landau two-fluid model.
Equation (\ref{0}) shows the  {\em entrainment effect},
i.e. the transport of the mass of both condensates by the superfluid motion
of each condensate, which arises due to the interaction between the quasiparticles with different isospin projections 
(Andreev \& Bashkin 1975, Vardanian \& Sedrakian 1981).  
In the neutron star cores the entrainment effect leads 
to the appearance of electric 
currents of protons driven by the neutron circulation, which become a source of internal axially symmetrical magnetic fields in pulsars (see Sedrakian \& Shahabasian 1991 and references therein).

Except for the early stages of pulsar evolution, the 
temperatures of the interiors of the stars,
 $T\sim10^8$ K, are much lower than the
critical temperatures for the superfluid phase transition 
($T_c\sim 10^9-10^{10}$ K), therefore the variation of the proton gap $\Delta_1(T=0)$ is along the density 
profile of the star. For typical parameters of 
the outer core of a a neutron star, 
the protons are type II superconductor (Baym, Pethick and Pines 1969). With increasing matter density, the attractive interaction between protons in the $^1S_0$ partial wave channel reduces, leading to a reducton of the proton gap. If the gap becomes sufficiently small,
the coherence length of the proton condensate, $\xi$, can become larger 
than the magnetic 
field penetration depth $\lambda$ 
(more precisely $\lambda/\xi\le 1/\sqrt{2}$) and the 
proton condensate will undergo a transition from a 
type II to a type I superconducting state.

The aim of this paper is the determination of the magnetic field distribution within a single neutron vortex line in the case when the proton condensate 
forms a type I superconductor. 
The present  problem bears  a certain resemblance to the problem of
the explanation of the anomalously large magnetic flux observed in the thermo-electrical experiments on the hollow superconducting cylinders. Our
calculation of the Gibbs free-energy parallels developments by
Aroutyunian \& Zharkov (1981) and Ginzburg \& Zharkov (1993) for a hollow superconducting sample in the presence of normal currents driven 
around a cylindrical cavity by the temperature gradients.

The paper is organized as follows. In Section 2 we set up the basic system of equations of macroscopic superfluid magnetohydrodynamics for the mixture
of neutron and proton condensates. In Section 3 these equations are applied  
to a single neutron vortex and the Gibbs free-energy is obtained for this case.
Section 4 shows that the minimum of the Gibbs potential
corresponds to a neutron vortex with a coaxial tube of normal protons confined 
within a finite radius $a$. The dependence of the radius of the tube and its 
homogeneous magnetic field on the density and the values of microscopic parameters is determined. Section 5 contains several concluding remarks.
 
\section{The Gibbs potential of a
superfluid neutron-proton mixture in a rotating 
system}

Consider the system of rotating superfluid neutron-proton mixture 
with a charge neutralizing background of relativistic electrons 
and a vanishing number of  quasiparticle excitations (zero temperature limit).
The kinetic part of the energy of neutron superfluid, 
which is by far the most dominant part of 
the energy of the system,  is minimized by a lattice
of neutron vortices. The superfluid executes a course-grained rigid body 
rotation with an angular velocity  ${\Omega}$, 
supported by a vortex lattice of density  
$n = 2\Omega/\kappa$ per unit area, where $\kappa$ is the 
quantum of circulation.
The entrainment effect does not changes this result to any 
considerable extent: the correction to the mass current of the neutrons is 
of the order of the ratio of proton to neutron density,
 which is typically of the order of  
$0.01 -0.05$. At the same time, the resulting magnetic energy density 
is lower by orders of magnitude  than the kinetic energy density.
Thus, the minimization of the total energy of 
the system can be carried out in two steps. 
In the first step, one minimizes the kinetic energy of 
rotating neutron superfluid. Once the neutron superflow pattern is fixed 
by minimization of the rotational energy, 
in the second step the thermodynamic corrections to the energy of 
the system are found by the minimization of the Gibbs free-energy associated 
with the unentrained part of the proton 
condensate (the motion of the entrained part follows  
that of neutron superfluid and is determined 
at the first step of the minimization procedure).
We  shall concentrate on the  second step of the minimization problem
under assumption that the proton liquid forms a type I superconductor.
A recent account of the case of type II superconductivity is given in 
Sedrakian \& Sedrakian (1995).

Taking account of the foregoing discussions, 
the thermodynamical Gibbs potential for a mixture 
of superfluid neutrons and superconducting protons can be written as
\be\label{1}
{\cal G}&=&{\cal F}  - {1\over c}\int{{\bmath A}\cdot {\bmath j}}_{12}~dV-
 {\bf\Omega} \times {\bmath M},
\ee
where ${\bmath j}_{12} $ is the entrainment current defined below 
(equation 8), 
${\bmath A}$ is the vector potential of the magnetic field, 
${\bf \Omega}$ is the angular velocity of the normal component, and the 
free-energy and the angular momentum of the system are 
\be\label{2}
{\cal F}&=&{\cal F}_{c} + \frac{1}{2}
\int\left(\rho_{11} v_1^2 + 2\rho_{12} \bmath v_1\cdot\bmath v_2 +\rho_{22} v_2^2 \right)dV\nonumber \\
&+&{1\over {8\pi}}  \int B^2dV -\frac{1}{2}\int\rho\left(\Omega\, r \right)^2 +{\bf\Omega}\times \bmath M,\\
\bmath M &=& \int \bmath r \times \left(\bmath p_1+\bmath p_2 \right)~dV ,
\ee
where $\rho$ is the total density of baryonic component.
The first term in equation (\ref{2}), ${\cal F}_{c}$, is the superfluid condensation energy.
The remaining terms are the sum of the kinetic and  the magnetic energy of the 
system. Here the gradient invariant superfluid velocities are 
defined relative to their rigid body rotation value ${\bf\Omega}
\times \bmath r$ 
\be 
\label{3}
\bmath v_1 &=& \frac{\hbar}{2m_1} {\bmath \nabla}\chi_1 - \frac{e}{m_1c}\bmath A ,\\
\label{4}
\bmath v_2 &=& \frac{\hbar}{2m_2} {\bmath \nabla}\chi_2 ,
\ee
where $m_{1,\, 2}$ denotes the bare mass, $\chi_{1,\, 2}$ the phase of the
superfluid order parameter in the rotating frame;
the isospin indecies $1$ and $2$ refer to protons and neutrons
respectively. As it can be seen from the relation (\ref{0}), 
the diagonal elements of the superfluid density matrix,
$\rho_{11}$ and $\rho_{22}$ are the densities of the unentrained parts of 
the proton and neutron condensates, respectively, while the off-diagonal elements
$\rho_{12} = \rho_{21}$ are the densities of entrained parts; these are 
related to the microscopic variables 
according to relations (Andreev \& Bashkin 1975; Vardanian \& Sedrakian 1981)
\be \label{5}
\rho_{12}={k\over {1+k}}\rho_1 ,\quad \rho_{11}={1\over {1+k}}\rho _1\, ,
\ee
where in the mean-field approximation the entrainment coefficient is
$k = (m_1^*-m_1)/m^*_1$, and $m_1^*$ and $m_1$ are the effective  
and the bare mass of protons, 
$\rho_1$ is the total density of the proton condensate, and the 
following sum rules hold
\be\label{6}
\rho_2 = \rho_{22} + \rho_{12}; \quad \rho_1=\rho_{11}+\rho_{12}; \quad
\rho=\rho_1+\rho_2 , 
\ee
where $\rho_2$ is the total density of the neutron condensate.

The magnetic induction 
${\bmath B} =\bmath\nabla\times {\bmath A}$ is determined from the Maxwell
equation, 
\be\label{7}
\bmath \nabla \times \,{\bmath B}=\frac{4\pi}{c}({\bmath j}_{11} + {\bmath j}_{12}), 
\ee
where the proton supercurrent is split into unentrained and entrained
parts, respectively,  
\be \label{8}
{\bmath j}_{11}={ e\over {m_1} }\rho_{11}{\bmath v}_1, \quad
{\bmath j}_{12}={ e\over {m_1} }\rho_{12}{\bmath v}_2.
\ee
On the scale of a single neutron vortex,
${\bmath j}_{12}$ is the entrainment current
which is generated by the neutron superfluid circulation in the neutron vortex, 
while  ${\bmath j}_{11}$ is the Meissner current appearing
due to the presence of unentrained proton condensate $(\rho_{11} \ne 0).$
Applying the curl operation to equation (\ref{7}) one finds the 
London equation for the case of type I proton superconductor 
\be\label{9} 
{\bmath B} + \lambda^{-2}\bmath \nabla \times \bmath \nabla \times  ~{\bmath B} 
= k{\bmath\phi}_0 \delta^{(2)}\left({\bmath r}-{\bmath r}_0\right),
\ee  
where $\phi_0=\pi\hbar c/e$ is the flux quantum and ${\bmath r}_0$ is 
the radius vector defining the position of neutron vortex in the lattice plane
and $\lambda$ is the magnetic field penetration depth,
\be 
       \lambda^2={ {m_1^2 c^2}\over {4\pi e^2\rho_{11} } }. \nonumber
\ee 
Further, a change of the origin of the free-energy
\be \label{10}
{\cal F'} = {\cal F} - {\cal F}_{c2} -\frac{1}{2}\int \rho_{22}\, v^2_2 ~dV
+\frac{1}{2}\int\rho\, \left(\Omega\, r\right)^2\, dV,
\ee 
can be made, since the neutron superfluid velocity, $\bmath v_{2}$, is essentially
the familiar solution of equation (\ref{4}) for a single neutron vortex, while ${\cal F}_{c2}$
- the condensation energy of neutrons - is an additive constant.
Finally, eliminating  the magnetic induction in favor of the vector 
potential from equation (\ref{3}) and denoting the condensation energy of 
protons by ${\cal F}_{c1}$, we find
\be\label{11} 
{\cal F'} &=& {\cal F}_{c1} + \int \Biggl[\frac{1}{8\pi\lambda^2}
\left(\frac{\phi_0}{2\pi}\bmath \nabla\chi_1-\bmath A\right)^2\nonumber \\
&+&\frac{1}{c}
\left(\frac{\phi_0}{2\pi}\bmath \nabla\chi_1-\bmath A \right)\cdot \bmath j_{12}
+\frac{1}{8\pi} \left(\bmath\nabla\times \bmath A\right)^2\Biggr]~dV.
\ee
Equations (\ref{1}), (\ref{7}), (\ref{9}) and (\ref{11}) form 
the basic system of equations of the problem.

\section{The distribution of equilibrium magnetic field within a neutron vortex}

Type I superconductors in the presence of imposed currents (or equivalently magnetic field)
break into successive domains of superconducting 
and normal regions; the geometry of 
the domains is determined by the symmetry of the problem.
Assume that the proton superconductor goes over to  the normal state
within a cylinder of a radius $a$, concentric with 
the neutron vortex, forming a tube of
a normal matter. (The cylindrical geometry of the tube  is dictated, 
in our case, by the symmetry of the neutron vortex).  
Let the volume inclosed within the radius $a$
be $V_<$, while that between the concentric cylinders 
$a$ and $b$, where $b$ is the outer radius of the neutron vortex, be $V_>$. 

Because the proton liquid is normal within the volume $V_<$, the superconducting
currents vanish in that region,  ${\bmath j}_{11} = {\bmath j}_{12} = 0$. 
The homogeneous magnetic field $B\equiv H_{cm}$ (we neglect for simplicity the paramagnetic 
effects for protons in the normal state)  is determined, in this region,
by the continuity condition $B_z =H_{cm}$ at the boundary $r=a$. 
The integration of equation (\ref{1}) over the volume $V_<$ is straightforward
\be\label{12}
{\cal G}^< = \frac{H_{cm}^2}{8\pi}\, V_<.
\ee
Decomposing the last term in equation (\ref{11}) in a sum of volume and surface 
integrals and using equation (\ref{7}), the Gibbs free-energy in the volume $V_>$ becomes 
\be\label{13} 
{\cal G}^> &=& -\frac{H_{cm}^2}{8\pi}\, V_<-\alpha_1 V_> \nonumber \\
&+&\int  \Biggl[\frac{\phi_0}{16\pi^2\lambda^2}\, 
\left(\frac{\phi_0}{2\pi}\bmath \nabla\chi_1-\bmath A\right)\cdot \bmath \nabla\chi_1\,
\nonumber \\
&+&\frac{1}{c}
\left(\frac{\phi_0}{2\pi}\bmath \nabla\chi_1-\bmath A \right)\cdot 
\bmath j_{12}-\frac{1}{2c}\, \bmath A\cdot \bmath j_{12}\Biggr]~dV_>,
\ee
where $\alpha_1$ is the condensation energy density of 
superconducting protons.

Let us turn to the determination of non-vanishing components of 
vectors ${\bmath A}$, $\bmath \nabla\chi_1$ and ${\bmath j}_{12}$. 
The  cylindrical symmetry of the problem  
implies that all unknown quantities are functions of the distance $r$ from 
the axis of the neutron vortex and the azimuthal angle $\varphi$. 
From equations. (\ref{3}) and (\ref{8}) the azimuthal components of 
vectors $\bmath \nabla\chi_1$ and  ${\bmath j}_{12}$ are  
\be\label{14}
({\bmath j}_{12})_{\varphi}=\frac{Q}{r} , \quad ({\nabla}\chi_1)_\varphi
=\frac{m}{r} ,
\ee
where $Q=(e\hbar/2m_1m_2)\rho_{12}$ and  $m$ is an integer number; the second 
relation follows form the quantization of the proton supercurrent circulation
around the cylindrical tube. 
The non-vanishing azimuthal component of the vector potential
$A_{\varphi}$ is  expressed in  terms of $B_z$ component of the magnetic induction using equation (\ref{7}): 
\be\label{15}
A_{\varphi} ={ {\hbar c} \over e} {m\over r} + { {4\pi\lambda^2Q}\over c}
{1\over r} + \lambda^2 { {dB_z}\over {dr} }.      
\ee
The boundary conditions for the magnetic inductions are $B_z(a) = H_{cm}$ and $B_z(b) = 0.$ 
From the first conditon it follows, particularly, that  the circulation of the vector 
${\bmath A}$ on the radius $a$ is 
\be\label{16}
      \oint {\bmath A}\cdot d{\bmath l}=\pi a^2H_{cm}.              
\ee
and therefore $A_\varphi (a)=(a/2)\, H_{cm}, $
or combining with equation (\ref{15})
\be\label{18}
  H_{cm}={ {m \phi_0}\over {\pi a^2} } + { {8\pi \lambda^2Q}\over{ca^2} }+
      {2\lambda^2\over a}{ {dB_z}\over {dr} }\Big| _{r=a} .  
\ee
The determination of $ A_\varphi$ function is accomplished by finding  
the function $B_z(r)$ from  equation (\ref{9}); 
the latter is a source-free in the 
region $r>a $. The solution with the boundary conditions above  is
\be\label{19}
 B_z(r)=\frac{\phi_0}{\pi a^2} \left(m+\frac{8\pi^2\lambda^2}{c\phi_0}\, Q\right) 
~\frac{{\cal N}(r)}{{\cal D}},
\ee
where 
\be\label{20}
{\cal N}(r) & = & 
    I_0\left( { b\over \lambda} \right) K_0\left( {r \over \lambda} \right)-
    K_0\left( { b\over \lambda} \right) I_0\left( {r \over \lambda} \right),  \\
\label{21}
 {\cal D} &= & I_0\left( b\over\lambda\right) K_2\left( {a\over \lambda} \right) -
      K_0\left( b\over\lambda\right)  I_2\left( {a\over \lambda} \right) ,
\ee
and $K$'s and $I$'s are the modified Bessel functions. Using this result, 
equation (\ref{18}) takes the form 
\be\label{22} 
 H_{cm}={ {\phi_0}\over {\pi a^2} } \left( m +
{ {8\pi\lambda^2} \over {{\cal L} \phi_0}}H_e(a)\right) {{\cal N}(a) \over {\cal D}}, 
\ee
where 
\be\label{23}
H_e(r)={ {4\pi}\over c}Q{\cal L},\quad {\cal L}={\rm ln}{b\over r}.    
\ee
Here $H_e(r)$ is the magnetic field intensity created by the 
entrainment currents. The
integrations in equation (\ref{13}) using equations (\ref{14}) and 
(\ref{22}) are straightforward, and the result for total Gibbs 
potential ${\cal G}=
{\cal G}^<+{\cal G}^>$ is 
\be\label{24}
{\cal G}&=&-\alpha_1 V_> +
{ {\phi_0^2} \over {8\pi^2a^2} } \Biggl\{ 
\left( m - p{ {\pi a^2 H_e(a)} \over {\phi_0} } \right) ^2\, \frac{{\cal N} (a)}{{\cal D}} \nonumber \\
&+&\frac{4\lambda^2}{a^2{\cal L}}
\left( { {\pi a^2 H_e(a)} \over {\phi_0} } \right)
\left( m - p{ {\pi a^2 H_e(a)} \over {\phi_0} } \right)
\frac{{\cal N} (a)}{{\cal D}}\nonumber \\ 
&+&q \left( { {\pi a^2 H_e(a)} \over {\phi_0} } \right)^2 \Biggr\} ,
\ee  
where $q=p_0-p$ and 
\be 
p&=&\frac{{\cal N}(a)}{{\cal D}} 
- { {2\lambda^2}\over {a^2 {\cal L}} }, \nonumber \\
p_0&=&{1\over {\cal L}^2 }\left( {b-a\over a} + {{(b-a)^2} \over {2a^2}} -
{\cal L} \right),
\ee
where $p_0$ is the limit of $p$ for $\lambda \to \infty$.
Note that equation (\ref{24}) has the proper ${\cal G}\to$ 0 behaviour when  
$\lambda\to \infty$  for a finite $(b-a)/b \ll 1$, i.e.  
of the Gibbs potential is measured from its value in the normal state 
(cf. Aroutyunian and Zharkov 1981). 

\section{The size of the normal proton tube within the neutron vortex}

As may be seen from the expression (\ref{24}), which contains two 
unknown parameters $m$ and $a$, the increase in the
energy of the system associated with the magnetic field is minimal if
one chooses  
\be\label{25} 
m=p{ {\pi a^2 H_e(a)} \over {\phi_0} }.               
\ee
Note that for the present problem (in contrast to the case considered
in Aroutyunian and Zharkov 1981) there are no energetic barriers for
the fulfillment of this condition, since for the superconducting protons in
the n-p-e phase of neutron stars hysteresis effects are absent.
Imposing the condition (\ref{25}) we find from equations
(\ref{22}) and (\ref{24}), respectively, 
\be \label{27}
H_e(a) = H_{cm}, \quad{\cal G}=-{\alpha_1 } V_> +q\pi a^2 \frac{H_{cm}^2}{8\pi}.
\ee
Note that the first condition coincides with the requirement $\bmath j_{12}=0$.
Using the definition $H_{cm}^2/8\pi = \alpha_1$ one finds 
\be
 H_{cm} = H_{t}(a) = 2\sqrt{2\pi \alpha_1}.
\ee
The relation determining  the radius $a$ follows by combining this expression 
with equation (\ref{23}):
\be\label{29}
a = b ~{\rm exp}\, \left(-\frac{2\sqrt{2\pi\alpha_1}}{H_0}\right),
\ee
where
\be\label{30}   
H_0={{k\phi_0}\over {2\pi \lambda^2} },\quad 
\alpha_1 =  \frac{\nu(\epsilon_F) \Delta_1^2}{4}, \quad
\nu= \frac{3^{1/3}\,m_1^{*\, 2/3}}{\pi^{4/3}\, \hbar^2}\, \rho_1^{1/3}. 
\ee
Here $\nu(\epsilon_F)$ is the density of states of protons on the Fermi-surface.

Table 1 gives the values of microscopic parameters of the proton superconductor 
for several values of total baryonic density. The dependence of the proton gap 
and the effective masses on the proton Fermi wave-vector, $k_{F1}$, are from 
Baldo {\it et al} (1992). The relation between $k_{F1}$ and the total baryonic 
density corresponds to the equation of state of Wiringa, Fiks \& 
Fabrocini (1988). 
The last column is the radius of the 
normal region following from equation (\ref{29}) for the 
neutron intervortex distance $b = 4 \times 10^{-3}$ cm, corresponding to the 
equilibrium rotation frequency of the Vela pulsar $\Omega = 70.6 $ s$^{-1}$.
If the proton gap eventually closes with increasing density, the radius of 
the normal region becomes $b$, i.e. the protons go over in the normal state in the whole volume of the vortex. 

%%%%%%%%%%%%%%%%%%%%%%%%%%%%%%%%%%%%%%%%%%%%%%%%%%%%%%%%%%%%%%%%%%%%%%%

\section{Concluding Remarks}

The present paper analyses the implications  of a transition 
of a proton superconductor
from type II to type I superconducting state for the 
magnetic structure of neutron vortex lattice  in the core of the neutron star, 
taking into account the entrainment effect.
The main result is that, owing to the entrainment 
effect, the type I proton 
superconductor breaks into  
successive normal superconducting regions. We analyzed 
the Gibbs potential in the case in 
which 
normal regions are of cylindrical shape, which  is most naturally 
implied by the  symmetry of the problem. We do not  exclude the possibility, however,
that the system may support 
a more complicated pattern of normal-superconducting domains, and 
those should be examined in future.

\setcounter{table}{0}
\begin{table*}
 \begin{minipage}{115mm}
        \caption{Parameter values and the size of the normal region in
                 type I proton superconductor}
\label{tab1}
        \begin{tabular}{cccccccc}
        \hline
$\rho$ & $k_{F1}$ & $m_1^*/m_1$ & $\Delta_1$ & $\xi$ & $\lambda$ & $H_0$ &$ a/b $ \\
$(\times 10^{14}$ g cm$^{-3}$) & (fm$^{-1})$&  &(MeV)& (fm) & (fm) & ($\times 10^{14}$ G)&   \\
\hline
7.91   & 0.85 & 0.69 & 0.3 & 54.2 & 41.58&  5.9 & 0.14\\
8.30  &  0.88 & 0.68 & 0.2 & 97.0 & 39.20 & 6.8 & 0.32 \\
8.56&  0.90   & 0.68  &  0.1 & 197.1 & 37.87&7.4 &  0.59  \\
\hline
   \end{tabular}
    \end{minipage}
                \end{table*}

The transition  from the
type I to the  type II superconducting state occurs at 
densities $\sim 3~ \rho_{sat}$.
These densities are indeed achieved in neutron star models 
with canonical masses $~\sim 1.4 ~M_{\odot}$
which are based on  moderately-soft equations of 
states (e. g. Wiringa { et al.} 
1988). Though the density range occupied by 
type I proton superconductor is not as broad as for 
a type II superconductor, the density profile of the star
at high densities is commonly quite flat, 
implying that the type I proton 
superconductor can occupy a considerable fraction of 
the volume of a superconducting core. 
The magnetic field moment, 
associated with the neutron vortex lattice, would be aligned
with the rotation vector unless the star is a non-axisymmetrical 
body at the instance of the nucleation of superconducting phase. 
Since this would require rotational velocities close to the mass-shedding 
limit, the nucleation of non-axisymmetrical field via this mechanism 
dose not appear to be a relistic possibility. The mean magnetic induction 
of a single neutron vortex is $\langle B\rangle 
\simeq H_e(a) \, (a/b)^2 \sim 10^{13}-10^{14}$ G for
densities listed in the table;  it drops, however,
to zero in the limit $a\to b$ when the entrainment currents vanish.
 
Present results on the ground state configuration of the type I proton 
superconductor can be used to address the problems of  dissipative motion 
of neutron vortex lattice, resulting in a effective coupling of the neutron superfluid  to the normal  electron liquid, heat generation via irreversible processes and other related problems.

Finally, we note that a situation similar to that discussed in the present
paper can be obtained in the metallic type I superconductors of cylindrical 
shape when a temperature gradient is applied between the axis and the boundary
of the sample. In principle, the effect of transition of the central part
of the cylinder to the normal state should be observable in this type of
experiments.

\section*{Acknowledgements}
DS is grateful to the P. N. Lebedev Physical Institute, Moscow, 
and DARC and CNRS, Observatoire de Paris, Meudon, for their hospitality.
AS gratefully acknowledges a research grant from the 
Max Kade Foundation,  NY.
GZ has been partially supported through the grant RFFI No 97-02-17545.

\end{document}